\documentclass[twocolumn,showpacs]{revtex4}

\usepackage{latexsym}
\usepackage{amssymb}
\usepackage{epsfig}
\usepackage{amsmath}
\usepackage{float}

\newcommand{\adotoa}{\ensuremath{{\cal H}}}

\newcommand{\grad}{\ensuremath{\vec{\nabla}}}
\newcommand{\Gammat}{\ensuremath{\tilde{\Gamma}}}

\newcommand{\Pb}{\ensuremath{\bar{P}}}
\newcommand{\rhob}{\ensuremath{\bar{\rho}}}

\newcommand{\phib}{\ensuremath{\bar{\phi}}}
\newcommand{\Zb}{\ensuremath{\bar{Z}}}

\newcommand{\kron}[2]{\ensuremath{\delta_{#1}^{\phantom{#1}#2}}}
\newcommand{\proj}[2]{\ensuremath{q_{#1}^{\phantom{#1}#2}}}
\newcommand{\pull}[2]{\ensuremath{h_{#1}^{\phantom{#1}#2}}}

\newcommand{\connh}{\ensuremath{\;^{(\tilde{q})}{\nabla}}}

\newcommand{\partialh}{\ensuremath{\tilde{\partial}}}
\newcommand{\Lie}[1]{\ensuremath{\underset{#1}{\mathcal{L}}}}

\newcommand{\nt}{\ensuremath{\tilde{n}}}
\newcommand{\qt}{\ensuremath{\tilde{q}}}
\newcommand{\et}{\ensuremath{\tilde{\epsilon}}}

\newcommand{\phit}{\ensuremath{\tilde{\phi}}}
\newcommand{\rhop}{\ensuremath{\rho^{(\phi)}}}
\newcommand{\rhobp}{\ensuremath{\bar{\rho}^{(\phi)}}}
\newcommand{\Pp}{\ensuremath{P^{(\phi)}}}
\newcommand{\Pbp}{\ensuremath{\bar{P}^{(\phi)}}}
\newcommand{\thetap}{\ensuremath{\theta^{(\phi)}}}

\newcommand{\be}{\begin{equation}}
\newcommand{\ee}{\end{equation}}
\newcommand{\bea}{\begin{eqnarray}}
\newcommand{\eea}{\end{eqnarray}}

\begin{document}

\title{Models of coupled dark matter to dark energy}

\author{
A.~Pourtsidou\footnote{E-mail: alkistis.pourtsidou@unibo.it}$^{a,b}$,
C.~Skordis\footnote{E-mail: skordis@nottingham.ac.uk}$^{c}$,
E.~J.~Copeland\footnote{E-mail: ed.copeland@nottingham.ac.uk}$^{c}$
}

\affiliation{
$^a$  Dipartimento di Fisica e Astronomia, Universit\`{a} di Bologna, viale Berti Pichat 6/2, 40127, Bologna, Italy 
\\
$^b$ Jodrell Bank Center for Astrophysics, University of Manchester, Manchester M13 9PL, UK
\\
$^c$School of Physics and Astronomy, University of Nottingham, University Park, Nottingham NG7 2RD, UK 
}

\begin{abstract}
We present three distinct types of models of dark energy in the form of a scalar field which is explicitly coupled to dark matter. 
Our construction draws from the pull-back formalism for fluids and generalises the fluid action to involve couplings to the scalar field.
We investigate the cosmology of each class of model both at the background and linearly perturbed level.
We choose a potential for the scalar field and a specific coupling function for each class of models 
and we compute the Cosmic Microwave Background and matter power spectra.
\end{abstract}


\maketitle

\section{Introduction}

During the last decade, observational cosmology has entered an era of unprecedented precision. Measurements of the
cosmic microwave background (CMB) (\cite{Komatsu:2010fb},\cite{Ade:2013zuv}), the Hubble constant ($H_0$)~\cite{Riess:2009pu}, the luminosity and distance at high redshift with 
supernovae Ia~\cite{Kowalski:2008ez}, and Baryon Acoustic Oscillations (BAO) surveys~\cite{Lampeitl:2009jq}, suggest that our Universe is currently undergoing a phase of 
accelerated expansion. The standard cosmological model, consisting of dark energy in the form of a cosmological constant ($\Lambda$) together with cold dark matter (CDM) fits all the available datasets extremely well, but it also suffers from two fundamental problems, namely the
\emph{fine-tuning} and \emph{coincidence} problems.  

However, there is plenty of margin left for alternative explanations for the nature of the dark sector, and many different approaches have been adopted (see~\cite{Copeland:2006wr} for a review). In particular, the
problems associated with the cosmological constant have led to a plethora of alternative theories, for example quintessence (\cite{Wetterich88, Ratra:1987rm, Wetterich:1994bg}), 
in which the DE component comes in the form of a scalar field evolving in time. An alternative approach is the modified
gravity scenario, which supports a completely different point of view, suggesting that our understanding of gravity through General
Relativity might not be applicable on the largest cosmological scales, i.e. that dark energy is due to a modification of gravity (see~\cite{CliftonEtAl2011} for a review).

Given that the precise nature of the two dark sectors is at present unknown, it may be that dark matter and dark energy have non-zero couplings to each other.
This possibility, which can be used to alleviate the coincidence problem, has been investigated in a number of cases in the past. Traditionally dark energy has been modelled as a scalar field $\phi$. If this field is uncoupled then
in order to be a viable model of dark energy (the primary property of which is to provide for cosmic acceleration) this field must today have negative pressure.
This is equivalent to saying that it's energy density is potential dominated. If however, this field is coupled to dark matter, then this may not be necessarily true. Indeed
the coupling to dark matter may also induce an effective negative pressure so that cosmic acceleration occurs even if the uncoupled field does not by itself have this property.
The Lagrangian of such a coupled system can be written as
\be
{\cal L} = -\frac{1}{2}\partial^\mu \phi \partial_\mu \phi -V(\phi) - m(\phi)\bar{\psi}\psi + {\cal L}_{\rm kin}[\psi],
\ee where $m(\phi)$ is the mass of the matter fields $\psi$. If we define the coupling current $J_\mu$ as
\be
J_\mu = - \frac{\partial {\rm ln} m(\phi)}{\partial \phi}\rho_c \partial_\mu \phi,
\ee 
where $\rho_c$ is the matter energy density, then the Bianchi identities can be written as
\be
\nabla_\nu T^\nu_{(\phi) \, \mu} = - J_\mu.
\ee
If $\phi$ is coupled to CDM only, we also have 
\be
\nabla_\nu T^\nu_{(c) \, \mu} = J_\mu,
\ee so that the total stress-energy momentum tensor of the dark components is conserved.
Models of this type have been thoroughly investigated. 
Amendola~\cite{Amendola:1999er} studied such a coupled quintessence (CQ) model assuming an exponential potential for $V(\phi)$ and a coupling of $\phi$ to the matter sector of the form $m(\phi) = \exp(\beta_c \phi)$, and investigated its cosmological consequences. He showed that the system could approach scaling solutions with $\Omega_\phi \simeq 0.7$ with an associated accelerated expansion and he constrained the coupling constant ($\beta_c$) using CMB data.

As the form of the coupling is chosen phenomenologically, there are a rich selection of papers investigating various interacting dark energy models and their cosmological implications, for example the effects of coupling on the Cosmic Microwave Background (CMB) and matter power spectra, on Supernovae, the growth of structure, non-linear perturbations and N-body simulations (see, e.g. \cite{Billyard:2000bh, Zimdahl:2001ar, Farrar:2003uw, Matarrese:2003tn, Amendola:2003wa, Maccio:2003yk, Amendola:2004ew, Amendola:2006dg, Guo:2007zk, Koivisto:2005nr, Lee:2006za, Wang:2006qw, Mainini:2007ft, Pettorino:2008ez, Xia:2009zzb, Chimento:2003iea, Olivares:2005tb, Sadjadi:2006qp, Brookfield:2007au, Boehmer:2008av, CalderaCabral:2009ja, He:2008tn, Quartin:2008px, Valiviita:2009nu, Pereira:2008at, Bean:2008ac, Gavela:2009cy, Tarrant:2011qe, Baldi:2008ay, Marulli:2011jk, Salvatelli:2013wra}). 
A different class of coupled dark energy models are those for which the coupling is to
neutrinos rather than CDM (\cite{Fardon:2003eh, Amendola:2007yx, Brookfield:2005bz, Mota:2008nj, Takahashi:2005kw}), more precisely by making the neutrino mass $\phi$ dependent. Several studies have also addressed the appearance of instabilities in coupled models (e.g. \cite{Valiviita:2008iv, Jackson:2009mz, Afshordi:2005ym, Bean:2007ny, Clemson:2011an, Takahashi:2006jt}).

In order to make further progress at the phenomenological level, it is desirable to construct general models of coupled dark energy. A natural question that arises is whether
the models considered so far saturate the possibilities. In other words, what is the most general phenomenological model one can construct? As we will show in this article,
the models considered so far are only one small subset in the space of possibilities. The common feature of those models is a coupling that involves the energy density (and
sometimes the pressure) of the dark fluid (whether CDM or neutrinos). However, it is also possible to couple to the velocity of the fluid as we shall show below (note that the consequences of a velocity coupling associated with dark matter scattering elastically with dark energy, leading to pure momentum transfer, were investigated in \cite{Simpson:2010vh}).

Our investigation highlights three classes of coupled models. The first two involve both energy and momentum transfer between the two components of the dark sector (albeit with distinctively different coupling mechanisms), while the third is identified as a pure momentum transfer model.
 
In Section II we give a short overview of the pull-back formalism for a fluid as is used in GR in order to construct an action from which the dynamics of the fluid are derived. We use this formalism to construct three distinct general classes (Types) of coupled dark energy models in Section III, and we derive the field equations of motion. In Section IV we study the background cosmology for the three Types of models. In Section V we derive the perturbed cosmological equations and investigate the observational implications on the CMB and matter power spectra for three specific cases (one for each Type). We conclude in Section VI.

\section{The pull-back formalism for fluids}
The coupled models of dark energy investigated in the past introduce the phenomenological coupling by modifying the field equations. However, it may be desirable to
introduce the coupling at the level of the action. The reason is that when building theories, using an action principle is in most cases more intuitive. Although we are interested in
phenomenological models of fluids for the case of CDM, rather than the actual fundamental field that may be CDM, using an action principle can provide better insight as to
how such couplings may emerge. We shall return to this further below.

We need a description of fluids at the level of the action. Fortunately such a description has already been formulated and is called the fluid {\it pull-back formalism}. 
 The fluid pull-back (FPB) formalism is a way to construct an action from which
 the dynamics of the fluid are derived.  It was formulated independently by 
 Kijowski, Sm{\'o}lski and G{\'o}rnicka~\cite{KijowskiSmolskiGornicka1990}, 
by Brown~\cite{Brown1993} and by  Comer and Langlois~\cite{ComerLanglois1993,ComerLanglois1994}, 
 building on earlier work by Taub~\cite{Taub1954} and Carter~\cite{Carter1973}.
The reader is referred to the review of Andersson and Comer~\cite{AnderssonComer2006} for further study.
 Before embarking on describing the three coupled models of dark energy, we first 
give a short overview of the pull-back formalism for a fluid as is used in GR.

One of the assumptions regarding perfect fluids is that collisions do not occur. Or put differently, the time taken to traverse a distance equal to the 
mean free path of the particles comprising the fluid is much larger than the
time for which our description is supposed to hold. In the case of cosmology, the mean free path of such fluids corresponds to times longer than many Hubble times.
The absence of collisions means that the trajectories of particles, called the worldlines, do not intersect. Furthermore, if the distribution of particles is continuous (as it should be for the 
fluid description to hold) then the space of worldlines forms a three-dimensional manifold $\cal{F}$. We may then introduce coordinates on this manifold which we denote as $X^I$ with
$I = 1\ldots 3$. Each point in $\cal{F}$ denotes a unique worldline associated with a distinct particle, meaning that the $X^I$ coordinates may be thought as particle labels.
Since the worldlines cannot intersect then there can be one and only one particle associated with each point in $\cal{F}$. 

The space $\cal{F}$ is time-less, i.e. each particle has its own 
specific label $X^I$ for all times. 
If this were not the case and it was possible for $X^I$ to change as time flows then
 it would also be possible for the worldlines to intersect. This means that the fluid description using the space of
 worldlines is the relativistic analogue to the Lagrangian
coordinate system in Newtonian mechanics, i.e. the system of coordinates which follows the particles as they move through space.

\subsection{The $3+1$ decomposition}
In order to describe the fluid using the manifold ${\cal F}$ we need to embed it in spacetime ${\cal M}$. Since ${\cal F}$ is three-dimensional it is in
one-to-one correspondence with a three-dimensional spacelike surface ${\cal S}_t$ at some particular time $t$. 
However, we have an infinite number of surfaces ${\cal S}_t$, one for each $t$. 

\begin{figure} [H]
\epsfig{file=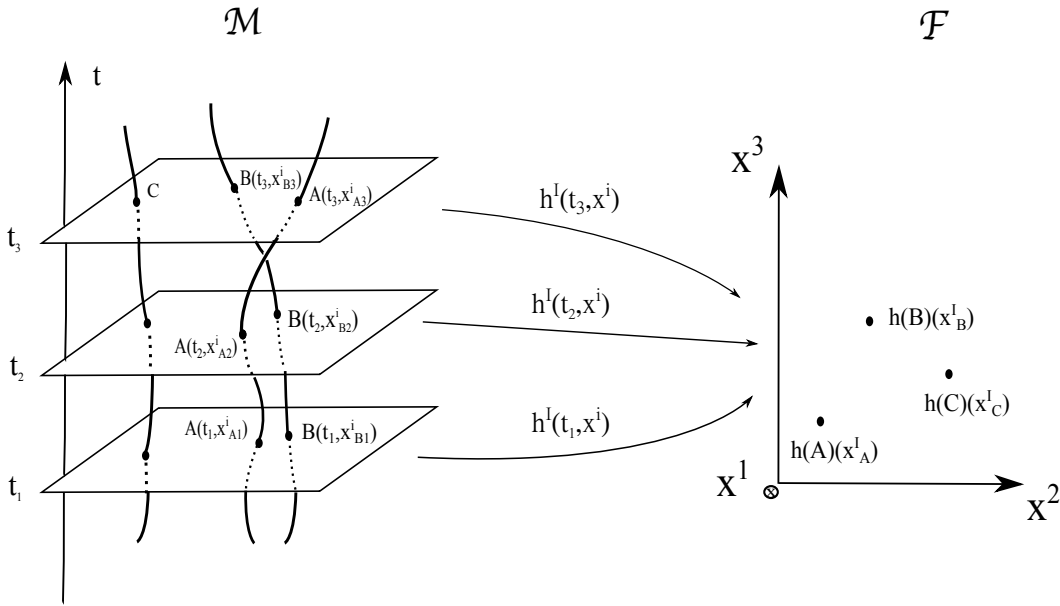,width=140mm,height=80mm}
\caption{The embedding of the three-dimensional (timeless) matter space ${\cal F}$ to fluid-element worldlines in space-time ${\cal M}$. The maps $ h^I(t,x^i)$ map the same point at coordinates $X^I$ to
a point $x^i$ in ${\cal M}$. The motion of particles in space-time is constructed by stacking all mappings for all times $t$.}
\label{figure_manifold}
\end{figure}

Consider a surface ${\cal S}_0$ at time $t_0$ and a timelike vector field $N^\mu$ which can be chosen to be normal to  ${\cal S}_0$.
By using the integral curves of $N^\mu$ parameterized by $t$, we can then foliate ${\cal M}$ by associating a surface ${\cal S}_t$ normal to $N^\mu$ for each $t$.
What remains is to associate each  ${\cal S}_t$ to  ${\cal F}$. This is done with the help of a map $h^I : {\cal S}_t \rightarrow {\cal F}$, where the index $I$ is to take into account the fact that
the two spaces involved are three-dimensional. If $x^\mu = \{t, x^i\}$ (with $i = 1\ldots 3$) are coordinates on ${\cal M}$, then under the map $h^I$ we have that $X^I = h^I(t,x^i)$. In other words, we have an infinite number of maps
labelled by $t$, which map the points of each ${\cal S}_t$ to  ${\cal F}$. For each $t$ the mapping is one-to-one. 
Although the points in ${\cal F}$ are fixed, motion is perceived through the embedding of ${\cal F}$ in ${\cal M}$. In other words, the maps $ h^I(t,x^i)$ map the same point at coordinates $X^I$ to
a point $x^i$ in ${\cal M}$ and by stacking together all mappings for all $t$ in a continuous way we have the appearance of motion of the particles in spacetime. This is shown in Fig.~\ref{figure_manifold}.
From the ${\cal M}$ perspective, the maps $ h^I(t,x^i)$ form a set of three scalar fields.
\onecolumngrid
\twocolumngrid

Now even at fixed $t$ the choice of ${\cal S}_t$ is not unique but depends on the choice of a timelike vector field $N^\mu$ to which  ${\cal S}_t$ is normal to.  There is one particular choice for $N^\mu$ which is
the $4$-velocity of the fluid particles, $u^\mu$, and without loss of generality we shall make this choice hence forth. 

The fluid velocity obeys  $u^\mu u_\mu = -1$ and we can use it to define a metric $q_{\mu\nu}$ on the hypersurface ${\cal S}_t$ as
\begin{equation}
q_{\mu\nu} = u_\mu u_\nu + g_{\mu\nu}.
\label{eq_q}
\end{equation}
The tensor $q_{\mu\nu}$ is also a projector, i.e. it obeys $q_{\mu\alpha} q^{\alpha}_{\phantom{\alpha}\nu} = q_{\mu\nu}$. Furthermore we have that
$q_{\mu\alpha} q^{\alpha\nu} = q_{\mu}^{\phantom{\mu}\nu}  = u_\mu u^\nu + \kron{\mu}{\nu}$
  and $q_{\mu}^{\phantom{\mu}\mu} = 3$. Note that indices are raised and lowered using the metric $g_{\mu \nu}$.

Any tensor can be projected onto parts parallel to $u^\mu$ and parts orthogonal to it via 
$q_{\mu}^{\phantom{\mu}\nu}$. For instance, consider the completely anti-symmetric tensor $\epsilon_{\mu\nu\alpha\beta}$ defined by $\epsilon_{\mu\nu\alpha\beta}= \sqrt{-g}\, \eta_{\mu\nu\alpha\beta}$ where
$g$ is the metric determinant and $\eta_{\mu\nu\alpha\beta}$ is the Levi-Civita anti-symmetric tensor density so that $\eta_{0123}=-1$ in all coordinate systems.
 We find $\epsilon_{\mu\nu\alpha\beta} =  - u_\mu u^\rho \epsilon_{\rho\nu\alpha\beta} +   q_{\mu}^{\phantom{\mu}\rho} \epsilon_{\rho\nu\alpha\beta}$
and repeating the procedure on the 2nd term (and so on) we finally get
 \begin{equation}
\epsilon_{\mu\nu\alpha\beta} = - u_\mu \epsilon_{\nu\alpha\beta} + u_\nu \epsilon_{\mu\alpha\beta} - u_\alpha  \epsilon_{\mu\nu\beta} +  u_\beta  \epsilon_{\mu\nu\alpha}, 
\end{equation}
where
\begin{equation}
\epsilon_{\mu\nu\lambda} = u^\rho \epsilon_{\rho\mu\nu\lambda}.
\label{epsilon_u}
\end{equation}
The last relation can be inverted to give
\begin{equation}
  u^\rho = \frac{1}{6} \epsilon_{\mu\nu\lambda} \epsilon^{\mu\nu\lambda\rho}.
\label{u_epsilon}
\end{equation}
The tensor $\epsilon_{\mu\nu\lambda}$ is completely antisymmetric and is related to the $3$-dimensional Levi-Civita tensor density
 $\delta_{ijk}$ as $\epsilon_{ijk} = \sqrt{q} \, \delta_{ijk}$ where $q$ is the $3$-metric determinant and $\delta_{123}=1$ in all coordinate systems.

Nothing discussed so far permits us to uniquely construct the tensors $u^\mu$, $q_{\mu\nu}$ and $\epsilon_{\mu\nu\lambda}$. Their construction involves knowledge about the fluid and as
we shall see further below, the fundamental variable from which these tensors are derived is the dual fluid number density tensor on ${\cal F}$, $\nt_{IJK}$.  It is therefore
not allowed  to use the relations (\ref{epsilon_u}) and (\ref{u_epsilon}) to obtain the variations $\delta u^\rho$ and $\delta \epsilon_{\mu\nu\lambda}$.

\subsection{Pullbacks and pushforwards for metrics, volume forms and connections}
So far we have considered tensors on the spacetime manifold ${\cal M}$. However, we also have tensors defined on the fluid manifold ${\cal F}$. For instance, the fluid manifold may also have a metric
$\qt_{IJ}$, or a completely anti-symmetric tensor $\et_{IJK} = \sqrt{\qt} \, \delta_{IJK}$ playing the role of a volume form, where $\qt$ is the determinant of $\qt_{IJ}$.

Tensors on ${\cal F}$ are related to tensors on ${\cal S}_t$ and by extension along the integral curves of $u^\mu$ to the whole of ${\cal M}$. This is achieved through the maps $h^I(t,x^i)$ via the pull-back and push-forward
operations (hence the name for this formalism). More specifically we can pull-back covariant vector fields (forms) from ${\cal F}$ to ${\cal M}$ and push-forward contravariant vector fields from ${\cal M}$ to ${\cal F}$
with the help of
\begin{equation}
 \pull{\mu}{I} = \nabla_\mu h^I.
\label{pullback}
\end{equation}
This is remiscent of coordinate transformations, in fact if ${\cal M}$ and ${\cal F}$ had the same dimensionality and the map was invertible then it reduces exactly to a coordinate transformation. 
With the help of (\ref{pullback}) we pull-back a form from $\tilde{v}_I$ on ${\cal F}$ to a form $v_\mu$ on  ${\cal M}$ as
\begin{equation}
  v_\mu = \pull{\mu}{I} \tilde{v}_I
\end{equation}
and push-forward a vector field $v^\mu$ from ${\cal M}$ to a vector field $\tilde{v}^I$ on ${\cal F}$  as
\begin{equation}
\tilde{v}^I = \pull{\mu}{I}  v^\mu.
\end{equation}

We can apply the formalism above to $\qt_{IJ}$ and $\et_{IJK}$ and relate them to $q_{\mu\nu}$ and $\epsilon_{\alpha\mu\nu}$ as
\begin{equation}
q_{\mu\nu} =  \pull{\mu}{I} \pull{\nu}{J} \qt_{IJ}
\end{equation}
and
\begin{equation}
\epsilon_{\alpha\mu\nu} =  \pull{\alpha}{I} \pull{\mu}{J} \pull{\nu}{K}  \et_{IJK}
\end{equation}
respectively. Clearly, a geometry on ${\cal F}$ induces a geometry on ${\cal S}_t$.

Let's now make things more interesting. We consider connections.
Given $g_{\mu\nu}$ we define the connection $\nabla_\mu$, s.t. $\nabla_\mu g_{\alpha\beta}= 0$,  and 
 associated Christoffel symbol $\Gamma^\alpha_{\mu\nu}$. Likewise on ${\cal F}$ we 
 define a connection $\connh_I$ with associated Christoffel symbol $\Gammat^K_{IJ}$.
The pull-back maps (\ref{pullback}) can then be used to relate the action of the two connections on forms on each corresponding space.
In particular for scalars we have
\begin{equation}
 \nabla_\mu \phi = \pull{\mu}{I} \connh_I \tilde{\phi}
\end{equation}
and on a form
\begin{equation}
 \nabla_\mu T_\nu = \pull{\mu}{I}  \pull{\nu}{J} \connh_I \tilde{T}_J.
\end{equation}
The Christoffel symbols are also related. From the above equation a straightforward calculation gives
\begin{equation}
 \pull{\lambda}{K} \Gamma^\lambda_{\mu\nu}  = \pull{\mu}{I}  \pull{\nu}{J} \tilde{\Gamma}^K_{IJ} +  \partial_\mu  \pull{\nu}{K}.
\label{conn_trans}
\end{equation}
The above relation is a generalization of the usual transformation law of Christoffel symbols under coordinate transformations, only here it is valid even if the dimensionality of the two spaces involved is
different.

\subsection{The relativistic fluid}

\subsubsection{Kinematical setup}
We can now proceed to fluids. Consider the fluid space ${\cal F}$ whose points denote particles for all time (worldlines). 
The density of points of the fluid space should then be related to the particle number density in some sense. 
Indeed, we can define a tensor  $\nt_{IJK}$ called the {\it dual} number density, which measures the number of particles within a region.
The total number of particles is then given by
\begin{equation}
 N =  \int \nt_{IJK} dX^I \wedge dX^J \wedge dX^K.
\end{equation}
The corresponding tensor in spacetime is given by the pull-back of  $\nt_{IJK}$
\begin{equation}
 n_{\mu\nu\lambda} =  \nt_{IJK}  \pull{\mu}{I} \pull{\nu}{J} \pull{\lambda}{K}
\label{n_pull}
\end{equation}
from which, in turn, we can get the particle number density $n$ as
\begin{equation}
n = \frac{1}{6} n^{\mu\nu\lambda} n_{\mu\nu\lambda}.
\end{equation}
 The number of particles can be obtained directly from spacetime as $N = \int n_{\mu\nu\lambda} dx^\mu \wedge dx^\nu \wedge dx^\lambda$.

We further define a number density current $n^\mu$ as the dual of $n_{\mu\nu\lambda}$ as
\begin{equation}
  n^\rho = \frac{1}{6} n_{\mu\nu\lambda} \epsilon^{\mu\nu\lambda\rho}
\label{n_mu}
\end{equation}
so that $n^2 = - n_\mu n^\mu$.
 Thus $n^\mu$ is timelike, and normalizing it defines the fluid velocity $u^\mu$ as
\begin{equation}
u^\rho = \frac{n^\rho}{n} = \frac{1}{6n} n_{\mu\nu\lambda} \epsilon^{\mu\nu\lambda\rho}.
\label{u_n}
\end{equation}
The push-forward of $n^\mu$ on the matter space vanishes, i.e. $\nt^I = \pull{\mu}{I} n^\mu = 0$.
The (\ref{n_mu}) relation can be inverted so that
\begin{equation}
n_{\mu\nu\lambda} = n^\rho \epsilon_{\rho\mu\nu\lambda}.
\label{n_three}
\end{equation}
Eq. (\ref{u_n}) uniquely defines the fluid velocity in terms of the fluid number density $n$ and dual number density $n_{\mu\nu\lambda}$. Having obtained $u^\mu$ via  (\ref{u_n}) we can then use (\ref{eq_q})
 to get the three metric $q_{\mu\nu}$ (and then push-forward $q^{\mu\nu}$ to get $\qt^{IJ}$ on the matter space). We can also obtain $\et_{IJK} = \frac{1}{n}\nt_{IJK}$
and pull-back it to spacetime to get $\epsilon_{\mu\nu\lambda}$. Alternatively 
the tensor $\epsilon_{\mu\nu\lambda}$ can also be obtained via (\ref{epsilon_u}) once $u^\mu$ is defined by  (\ref{u_n}).

\subsubsection{The adiabatic fluid action and variations}
We are now ready to discuss the dynamics of fluids as they are derived from an action. The dynamical variables are the spacetime metric and the fluid coordinates $X^I$ on ${\cal F}$
which from the spacetime perspective are the maps $h^I(x^\mu)$ which may be considered as three scalar fields. 

As we have seen in the last subsection, the quantity of interest
which depends on  $X^I$  is the number density $\nt$ (or $n$ from the spacetime perspective). 
Therefore, the action for General Relativity coupled to an adiabatic fluid is taken to be
\begin{equation}
S = \frac{1}{16\pi G} \int d^4x \sqrt{-g} R - \int d^4x \sqrt{-g} f(n),
\end{equation}
where $f(n)$ is an arbitrary function. We shall further see  below that the equation of state and its speed of sound 
is determined entirely by the form of $f$. Specifically, for pressureless matter $f$ is proportional to the number density: $f = f_0 n$.

The field equations are obtained via a variational principle. For brevity let us define
\begin{equation}
\delta X^I = y^I,
\end{equation}
a definition which will be useful when we discuss cosmological perturbation theory. Furthermore, we will find it useful to work with $y^\mu$ which is 
related to  $y^I$ by $y^I = \pull{\mu}{I}  y^\mu$.

Since $\delta f = \frac{df}{dn} \delta n$ we need the variation of the number density $n$. To find that we need the variation of $n_{\mu\nu\lambda}$ which from (\ref{n_pull}) is
given by
\begin{eqnarray}
\delta n_{\mu\nu\lambda} &=& 
(\partialh_L\tilde{n}_{IJK}) \pull{\mu}{I} \pull{\nu}{J} \pull{\lambda}{K} y^L
 + \tilde{n}_{IJK} \bigg[  
 \pull{\nu}{J} \pull{\lambda}{K} \nabla_\mu y^I 
\nonumber 
\\
&&
+ \pull{\mu}{I} \pull{\lambda}{K} \nabla_\nu y^J 
+ \pull{\mu}{I} \pull{\nu}{J} \nabla_\lambda y^K
 \bigg].
\label{eq_delta_n_1}
\end{eqnarray}
The  above expression may look non-covariant as it contains partial derivatives. But as we now show, it is. Since
 $\nabla_\mu y^I = \pull{\nu}{I}  \nabla_\mu y^\nu  +  y^\nu\nabla_\mu \pull{\nu}{I} $
and using (\ref{conn_trans}) we find
\begin{eqnarray}
&&
 (\partialh_L\nt_{IJK})  \pull{\mu}{I} \pull{\nu}{J} \pull{\lambda}{K} y^L  
  =  y^\rho\bigg[ \nabla_\rho n_{\mu\nu\lambda}  
\nonumber 
\\
&&
 \ \ \ \ 
- \nt_{PJK} \pull{\nu}{J} \pull{\lambda}{K}  \nabla_\mu\pull{\rho}{P}
 -\nt_{IPK} \pull{\mu}{I} \pull{\lambda}{K} \nabla_\nu\pull{\rho}{P}
\nonumber 
\\
&&
 \ \ \ \ 
 - \nt_{IJP} \pull{\mu}{I}\pull{\nu}{J}  \nabla_\lambda\pull{\rho}{P}
\bigg]. 
\label{eq_delta_n_2}
\end{eqnarray}
So combining (\ref{eq_delta_n_1}) and (\ref{eq_delta_n_2})  we get the simple relation
\begin{equation}
  \delta n_{\mu\nu\lambda} = \Lie{y}  n_{\mu\nu\lambda}, 
\label{delta_n_three}
\end{equation}
where $\Lie{y}$ is the Lie derivative along $y^\mu$.
 In fact the last relation is very useful and we shall use it hence forth.
To find $\delta n$ is now straightforward.  We use $\delta n = \frac{1}{12n} \delta (n^{\mu\nu\lambda} n_{\mu\nu\lambda})$ to get
\begin{equation}
  \delta n =  y^\mu \nabla_\mu n +  n \left[  \proj{\nu}{\mu} \nabla_\mu y^\nu + \frac{1}{2} q_{\mu\nu}  \delta g^{\mu\nu} \right].
\label{delta_n}
\end{equation}
Although, $\delta u^\mu$ is not needed to derive the field equations in the case of GR, we derive it here for completeness as
it will be needed when we couple the fluid to a scalar field. We use (\ref{u_n}) to get
\begin{equation}
\delta u^\rho = -\frac{1}{2} u^\rho u_\mu u_\nu \delta  g^{\mu\nu} + y^\mu \nabla_\mu u^\rho - u^\mu \proj{\nu}{\rho}  \nabla_\mu y^\nu.
\end{equation}

\subsubsection{The fluid equations}
Using (\ref{delta_n}) it is straightforward to find the fluid equations. The Einstein equations  $G_{\mu\nu} = 8\pi G T_{\mu\nu}$ are obtained as usual
by varying with the metric, and  the stress-energy tensor $T_{\mu\nu}$ is given by
\begin{equation}
T_{\mu\nu} = (\rho + P) u_\mu u_\nu +  P g_{\mu\nu},
\label{eq_T_fluid}
\end{equation}
where $\rho$ is the density and $P$ the pressure. From the variation of the fluid action and using  (\ref{delta_n}) we can match
 $\rho$ and $P$ to the fluid function $f(n)$ as
\begin{equation}
\rho =  f
\label{eq_rho_f}
\end{equation}
and
\begin{equation}
P =  n \frac{df}{dn} -f. 
\label{eq_P_f}
\end{equation}
Since we see that the speed of sound of the fluid is given by  
\begin{equation}
C_s^2 = \frac{\delta P}{\delta \rho} = C_a^2 = \frac{dP}{d\rho} =n \frac{f_{nn}}{f_n},
\end{equation}
where $f_n  = \frac{\partial f}{\partial n}$ and $f_{nn}  = \frac{\partial^2 f}{\partial n^2}$, 
it is clear that the fluid we have been considering so far is adiabatic. We shall use this notation for the rest of the paper
so that for a function $F(Y,Z,\ldots)$ we have $F_Y = \frac{\partial F}{\partial Y}$ and $F_{YZ}  = \frac{\partial^2 F}{\partial Y \partial Z}$ and so on.

Varying with respect to $X^I$ the field equations for the fluid are obtained as
\begin{eqnarray*}
  \frac{df}{dn}  \nabla_\mu n - \nabla_\nu\left(  n \frac{df}{dn}  \proj{\mu}{\nu}\right)  = 0.
\end{eqnarray*}
The above equation can be put in a more familiar form. We use (\ref{eq_rho_f}) and (\ref{eq_P_f}) as well as $ \frac{df}{dn} \nabla_\mu n = \nabla_\mu \rho$ 
and then project the resulting equation along $u^\mu$ and orthogonal to $u^\mu$ to get the energy conservation equation
\begin{equation}
 u^\nu \nabla_\nu \rho + (\rho+P) \nabla_\nu u^\nu = 0
\label{eq_rho_evolve}
\end{equation}
and momentum transfer equation
\begin{equation}
 \proj{\mu}{\nu}\nabla_\nu P + (\rho+P) u^\nu \nabla_\nu u_\mu = 0.
\end{equation}
Interestingly, the fluid coordinates $X^I$ and number density $n$ do not appear in the above equations and are no longer needed.

\section{Generic models for coupled fluids}
\subsection{Construction}
We now construct a model where the adiabatic fluid described  in the previous section is explicitly coupled to 
a scalar field $\phi$ which will play the role of dark energy.
The most general action formed from the fluid variable $X^I$ through the pull-back $\pull{\mu}{I}$, as well as the scalar field $\phi$ and its derivative 
\begin{equation}
\phi_\mu \equiv \nabla_\mu \phi
\end{equation}
 is 
\begin{equation}
S = \frac{1}{16\pi G} \int d^4x \sqrt{-g} R - \int d^4x \sqrt{-g}  L(\pull{\mu}{I},\phi,\phi_\mu). 
\end{equation}

We need the most general form for $L(\pull{\mu}{I},\phi,\phi_\mu)$.
Since we must have a tensor with $I$-indices to contract with $\pull{\mu}{I}$ and since $L$ must reduce to the GR case if $\phi$ is absent,
then the only dependence of $L$ on $\pull{\mu}{I}$ must come through $n_{\mu\nu\lambda}$ and therefore from (\ref{u_n}) also $u^\mu$. Therefore we can write 
\begin{equation}
L = L(n_{\mu\nu\lambda},u^\mu,\phi,\phi_\mu).
\end{equation}
Next we need to form invariants of the tensors $n_{\mu\nu\lambda}$, $u^\mu$, $\phi$ and $\nabla_\mu\phi$. Clearly we have as before
the invariant $n$ which once again plays the role of the fluid number density.  But we have also other invariants, namely
\begin{equation}
Y = \frac{1}{2} \phi_\mu \phi^\mu 
\end{equation}
which can be used to construct a kinetic term for $\phi$
 and 
\begin{equation}
Z = u^\mu \phi_\mu 
\end{equation}
which plays the role of a direct coupling of the fluid velocity to the gradient of the scalar field.
Since $n^{\mu\nu\lambda}$ is completely
antisymmetric any contraction of it with $g_{\mu\nu}$ or $\nabla_\mu\phi \nabla_\nu\phi$ vanishes. There may be combinations 
involving $2$nd derivatives of the metric, $X^I$ or $\phi$ which for miraculous reasons give 2nd order field equations as in the case 
of Horndeski theory, but we ignore this possibility for now.
Therefore our final functional form is
\begin{equation}
L = L(n,Y,Z,\phi).
\end{equation}
It is clear that ordinary GR with a quintessence field and a fluid is described by $L =  Y + V(\phi) + f(n)$ while k-essence is described by $L = F(Y,\phi) + f(n)$.

The above entity we have constructed is still fairly general. In particular we cannot split it into a scalar field part and a fluid part which are then
coupled, without further assumptions. We shall proceed to do that shortly. 

\subsection{Field equations for the generic fluid}
To get the field equations we proceed as in the case of ordinary relativistic fluid. 
We have $\delta L = L_n \delta n + L_\phi \delta \phi + L_Y \delta Y + L_Z \delta Z$ where
\begin{equation}
 \delta Y = \frac{1}{2} \phi_\mu \phi_\nu \delta g^{\mu\nu} + \phi^\mu \nabla_\mu \delta\phi
\end{equation}
and
\begin{equation}
 \delta Z = \phi_\mu \delta u^\mu + u^\mu \nabla_\mu \delta\phi.
\end{equation}
Varying with $\phi$ we find the scalar field equations as
\begin{equation}
\nabla_\mu\left( L_Y  \phi^\mu  +  L_Z u^\mu \right) - L_\phi = 0.
\label{eq_gen_phi}
\end{equation}
Varying with $g^{\mu\nu}$ we get the Einstein equations to be once again
$G_{\mu\nu} = 8\pi G T_{\mu\nu}$ where the total energy-momentum tensor is
\begin{equation}
 T_{\mu\nu} = L_Y  \phi_\mu  \phi_\nu + ( nL_n - Z L_Z ) u_\mu u_\nu  
 + ( n L_n - L) g_{\mu\nu}. 
\label{gen_T}
\end{equation}
Finally varying with $X^I$ we find the evolution equation for the number density $n$ as
\begin{eqnarray}
&&
-n  \nabla_\mu  L_n  +\nabla_\nu\left[ (Z L_Z -    n L_n)  u_\mu u^\nu + L_Z \phi_\mu  u^\nu \right] 
\nonumber
\\
&& \ \ \ \ \ \ \ \
 + L_Z \phi_\rho \nabla_\mu u^\rho 
\label{eq_gen_fluid}
=0.
\end{eqnarray} 
Contracting with $u^\mu$ we find the conservation law
\begin{equation} 
\nabla_\mu\left(n u^\mu\right) = \nabla_\mu n^\mu =0
\label{eq_cons_law}
\end{equation}
which when used in (\ref{eq_gen_fluid}) gives the purely spatial equation
\begin{eqnarray}
 && -nD_\mu L_n + L_Z D_\mu Z + \left(Z L_Z - nL_n\right) u^\nu \nabla_\nu u_\mu
\nonumber 
\\ && \ \ \ \ \ 
 + \phit_\mu \nabla_\nu\left( L_Z u^\nu\right) = 0,
\label{eq_gen_fluid_spat}
\end{eqnarray}
with $D_\mu =Ê q^\nu_{\; \mu} \nabla_\nu$ and $\phit_\mu = q^\nu_{\; \mu} \nabla_\nu \phi$.
Our system comprises of two scalars, $n$ and $\phi$ whose evolutions are completely determined via (\ref{eq_gen_phi}), (\ref{eq_cons_law})
and (\ref{eq_gen_fluid_spat}) provided the function $L$ is given. 

As we have already stressed, we cannot in general separate out a field and a fluid: we are dealing with a single entity. Problems such as this have been discussed in~\cite{Kunz:2007rk}. We may proceed, however, to do that under further assumptions. In particular we will consider three types of models. Type-1 models are those for which there is no $Z$ dependence and furthermore the function $L$ can be separated as $L = F(Y,\phi) + f(n,\phi)$. In both Type-2 and Type-3 models the function $L$ has $Z$ dependence and the difference between the two is how this $Z$ dependence appears. For Type-2, $L$ can
be separated as $L = F(Y,\phi) + f(n,Z)$ while for Type-3 the separation is as $L = F(Y,Z,\phi) + f(n)$. We now proceed to reduce the general equations above for each of the three types of models and in the process eliminate the variable $n$ and instead describe the fluid evolution in terms of an energy density $\rho$ and pressure $P$.

\subsubsection{Type-1 coupled dark matter to a dark energy scalar field}

Type-1 models, are classified via
\begin{equation}
L(n,Y,Z,\phi) = F(Y,\phi) + g(n) e^{\alpha(\phi)},
\end{equation} 
where we have set $f(n,\phi) =  g(n) e^{\alpha(\phi)}$. This later condition of separability of $f$ is not needed in order to simplify the field equations, but is needed in order to be able to solve them uniquely without resorting to the variable $n$.
These types of models describe a $K$-essence scalar field coupled to matter.
By further choosing $F=Y+V(\phi)$, we can also describe coupled quintessence models.

The fact that we can separate the function $L$ in this way allows us to
separate the general energy-momentum tensor (\ref{gen_T}) into 
an energy-momentum tensor for $\phi$ as
\be
T^{(\phi)}_{\mu\nu} = F_Y  \phi_\mu  \phi_\nu  - F g_{\mu\nu} 
\label{eq_T_phi_1}
\ee
and an energy-momentum tensor for the fluid given by (\ref{eq_T_fluid}) with
$\rho$ and $P$ identified as
\begin{equation}
\rho =    f 
\label{eq_rho_1}
\end{equation}
and
\begin{equation}
P = n f_n  -f
\label{eq_P_1}
\end{equation}
so that $\rho+ P = n f_n$.
The scalar field energy density $\rhop$ and pressure $\Pp$ are read-off from (\ref{eq_T_phi_1}) as
\begin{equation}
\rhop =  Z^2 F_Y +  F 
\label{eq_rhophi_1}
\end{equation}
and
\begin{equation}
\Pp =   \frac{1}{3} F_Y (Z^2 + 2 Y) - F. 
\label{eq_Pphi_1}
\end{equation}

We now proceed to simplify the field equations for this type of coupled models.
The scalar field equation (\ref{eq_gen_phi}) becomes
\begin{equation}
\nabla_\mu\left(F_Y  \phi^\mu  \right) -F_\phi = \rho \alpha_\phi,
\label{eq_phi_T1}
\end{equation}
while the fluid equation (\ref{eq_gen_fluid_spat}) gives
\begin{equation}
    D_\mu  P    + (\rho + P) u^\nu    \nabla_\nu  u_{\mu} =
 - \rho \alpha_\phi \phit_\mu.
\label{eq_fluid_spat_1}
\end{equation}
The evolution of the fluid density $\rho$ is found
from the conservation equation (\ref{eq_cons_law}) as
\begin{equation}
u^\mu  \nabla_\mu \rho + (\rho+P)\nabla_\mu u^\mu=
 Z \rho \alpha_\phi.
\end{equation}
Finally, let us calculate the coupling current $J_\mu = - \nabla_\nu T^\nu_{(\phi)\;\mu}$. From (\ref{eq_T_phi_1}) and using (\ref{eq_phi_T1}) we find
\begin{equation}
 J_\mu =  - \rho \alpha_\phi \phi_\mu.
\end{equation}

\subsubsection{Type-2 coupled dark matter to a dark energy scalar field}
Type-2 models are classified via
\begin{equation}
L(n,Y,Z,\phi) = F(Y,\phi) + f(n,Z).
\end{equation} 
In these type of models, the energy-momentum tensor of the scalar field is given
by (\ref{eq_T_phi_1}) as in the Type-1 case while the energy-momentum of the fluid is
again given by (\ref{eq_T_fluid}). However, unlike the Type-1 case,
the energy density $\rho$ of the fluid is now identified with
\begin{equation}
\rho = f - Z f_Z,
\label{eq_rho_2}
\end{equation}
while the pressure $P$ is still given by (\ref{eq_P_1}).
In the case that the scalar is coupled to CDM ($P=0$), the pressure equation (\ref{eq_P_1}) can be solved to give
 $f=nh(Z)$. Since we are only interested in the $P=0$ case  we shall use this functional form for $f$ in this type of coupled model.
 Furthermore, for reasons that will become clear, rather than using $h(Z)$ we introduce a new coupling function $\beta(Z)$ so that 
\begin{equation}
h(Z) = e^{\int dZ\frac{\beta}{1+Z\beta}}.
\end{equation}
Once again, since we have $F(Y,\phi)$ these types of models describe a $K$-essence scalar field coupled to matter
and by further choosing $F=Y+V(\phi)$,
 we can also reduce it to coupled quintessence models.

Let us now proceed to the field equations. The scalar field equation (\ref{eq_gen_phi}) for these type of models simplifies to
\begin{equation}
\nabla_\mu\left(F_Y \phi^\mu  \right)-F_\phi= 
- \nabla_\mu\left( \rho\beta u^\mu \right).
\label{eq_phi_2_a}
\end{equation}
By computing $u^\mu \nabla_\mu\rho$ using (\ref{eq_rho_2}) and then using  the conservation equation (\ref{eq_cons_law}) we find the evolution equation
for the energy density of the fluid as
\begin{equation}
 u^\mu \nabla_\mu \rho + \rho\nabla_\mu u^\mu = - Z\nabla_\mu\left(\rho\beta u^\mu\right).
\label{eq_phi_2_b}
\end{equation}
The last field equation is the momentum transfer equation for the fluid (\ref{eq_gen_fluid_spat}) which reduces for
this type of models to
\begin{eqnarray}
\rho  u^\nu \nabla_\nu  u_\mu = \nabla_\rho\left( \rho \beta u^\rho \right)\phit_\mu. 
\label{eq_phi_2_c}
\end{eqnarray}

The equations (\ref{eq_phi_2_a}), (\ref{eq_phi_2_b}) and (\ref{eq_phi_2_c}) still need some processing. In particular (\ref{eq_phi_2_a}) should be solved in terms of a $\ddot{\phi}$ term
 and should not contain  $u^\mu \nabla_\mu \rho$ terms and similarly for the other two equations.
Starting from (\ref{eq_phi_2_b}) we can rearrange it so that it be comes
\begin{equation}
 u^\mu \nabla_\mu \rho + \rho\nabla_\mu u^\mu = - \frac{Z\rho \beta_Z}{1+Z\beta}  u^\mu \nabla_\mu Z
\label{eq_phi_2_b2}
\end{equation}
from which we find
\begin{equation}
\nabla_\mu\left( \rho \beta u^\mu\right) = \frac{\rho \beta_Z}{1+Z\beta}  u^\mu\nabla_\mu Z.
\label{eq_x2}
\end{equation}
Using (\ref{eq_x2}) into (\ref{eq_phi_2_c}) we find
\begin{eqnarray}
 u^\nu \nabla_\nu  u_\mu = \frac{\beta_Z u^\nu\nabla_\nu Z}{1+Z\beta} \phit_\mu.
\label{eq_phi_2_c2}
\end{eqnarray}

Let us now proceed to the scalar equation (\ref{eq_phi_2_a}). We  need 
\begin{equation}
\square \phi = -u^\mu  \nabla_\mu Z +(u^\mu \nabla_\mu u^\nu) \phit_\nu + q^{\mu\nu} \nabla_\mu \phi_\nu
\end{equation}
and also $\phi^\mu \nabla_\mu Y$. Using  equation (\ref{eq_x2}) the latter gives
\begin{eqnarray}
\phi^\mu \nabla_\mu Y &=& 
 Z^2 u^\mu  \nabla_\mu Z 
- Z \phit^\mu \nabla_\mu Z
- Z \phi^\mu \phi^{\nu} \nabla_\mu u_\nu
\nonumber 
\\
&&
\ \ \ \
+\phi^\mu \phi^\nu \nabla_\mu \phit_\nu.
\end{eqnarray}

Then the scalar equation (\ref{eq_phi_2_a}) becomes
\begin{eqnarray}
&&
\left[ -F_Y + Z^2 F_{YY} +  \frac{ (F_Y \phit_\nu \phit^\nu + \rho ) \beta_Z}{1+Z\beta} \right] u^\mu  \nabla_\mu Z 
\nonumber 
\\
&&
\ \ \ \
 + F_{YY} \phi^\mu \phi^{\nu} \left(  \nabla_\mu \phit_\nu - Z \nabla_\mu u_\nu\right)
+ F_Y q^{\mu\nu} \nabla_\mu \phi_\nu
\nonumber 
\\
&&
\ \ \ \
 - Z F_{YY} \phit^\mu D_\mu Z
 +2Y F_{Y\phi} 
-F_\phi
= 0. 
\label{eq_phi_2_a2}
\end{eqnarray}
 Eq. (\ref{eq_phi_2_b2}), (\ref{eq_phi_2_a2}) and (\ref{eq_phi_2_c2}) are the final set of equations we need.
Finally, the coupling current $J_\mu$ is found to be
\begin{equation}
 J_\mu = \nabla_\nu\left(\rho\beta u^\nu \right) \phi_\mu. 
\end{equation}
Our choice to parameterize $h(Z)$ in terms of an integral over a new function $\beta(Z)$ now becomes clear through the simplicity of the above equations.

\subsubsection{Type-3 coupled dark matter to a dark energy scalar field}
Our third and final class of models is defined by
\begin{equation}
L(n,Y,Z,\phi) = F(Y,Z,\phi)  + f(n).
\end{equation}
This also allows us to separate the energy-momentum tensor for $\phi$ from the energy-momentum tensor of the fluid so that
\begin{equation}
 T^{(\phi)}_{\mu\nu} = F_Y \phi_\mu \phi_\nu  - F g_{\mu\nu}- Z F_Z   u_\mu u_\nu  
\end{equation}
giving the field energy density as
\begin{equation}
\rhop =  Z^2 F_Y- Z F_Z +  F 
\end{equation}
while the pressure is as before given by (\ref{eq_Pphi_1}).
The energy-momentum tensor of the fluid is given again by (\ref{eq_T_fluid})
where the energy density and pressure are identified with $f$ using (\ref{eq_rho_f}) and
(\ref{eq_P_f}).

Proceeding to the field equations, the scalar field equation (\ref{eq_gen_phi}) simplifies to
\begin{equation}
\nabla_\mu\left(F_Y  \phi^\mu +F_Z u^\mu \right)-F_\phi= 0.
\end{equation}
Using the conservation equation (\ref{eq_cons_law}) we find that the energy-density $\rho$ of the fluid evolves as in the standard case with (\ref{eq_rho_evolve}) while
the momentum-transfer equation (\ref{eq_gen_fluid_spat}) becomes
\begin{eqnarray}
&& D_{\nu}  P + \left(\rho + P - ZF_Z\right) u^\beta    \nabla_{\beta}  u_{\mu} 
=
 \nabla_\beta \left( F_Z  u^\beta\right) \phit_\mu
\nonumber 
\\ && \ \ \ \ 
    +F_Z D_\mu Z. 
 \label{dP_type3}
\end{eqnarray}
Finally, the coupling current $J_\mu$ is found to be
\begin{equation}
 J_\mu = \nabla_\nu\left( F_Z u^\nu \right) \phit_\mu +F_Z D_\mu  Z  + Z  F_Z  u^\nu \nabla_\nu u_\mu.
\label{J_type3}
\end{equation}
Having the required field equations at hand we now proceed to apply them to cosmology. For simplicity we shall only consider the case of
coupled quintessence, although in a follow up paper we will discuss more general cases. 

\section{Cosmology}

\subsection{Cosmological equations}
To study cosmology and in particular the observational effects of the coupled models on the Cosmic Microwave Background and Large Scale Structure,
 we consider linear perturbations about the FRW spacetime. As is common, we choose the synchronous gauge so that the metric  is
\begin{eqnarray}
ds^2 &=& -a^2d\tau^2 + a^2 \left[ (1 + \frac{1}{3}h)\gamma_{ij} + D_{ij}\nu\right] dx^i dx^j,
\ \ 
\label{perturbed_metric}
\end{eqnarray}
 where $\tau$ is the conformal time, $\grad_k$ is the covariant derivative associated with $\gamma_{ij}$, i.e. $\grad_k \gamma_{ij} = 0$
and $D_{ij}$ is the traceless derivative operator $D_{ij} = \grad_i \grad_j -\frac{1}{3} \grad^2 \gamma_{ij}$.
The unit-timelike vector field $u^\mu$ is perturbed as
\begin{eqnarray}
u_\mu &=& a( 1, \grad_i \theta).
\label{u_mu_theta}
\end{eqnarray}
 A dot is derivative with respect to conformal time: $\dot{A}=\frac{dA}{d\tau}$.
We will use a "bar" over a variable to denote it's FRW reduction, i.e. $\rhob$ is the FRW background energy density for the fluid.

The Einstein equations for all types of coupling are as usual given by the Friedman equation 
\begin{equation}
3\adotoa^2 = 8\pi G a^2 \sum_i \rhob^{(i)}
\end{equation}
at the background level, where ${\cal H} \equiv \frac{\dot{a}}{a}$. At the perturbed level we have
\begin{equation}
\dot{h} = \frac{1}{\adotoa} \left[8\pi G a^2 \sum_i \rhob^{(i)} \delta^{(i)}  + 2 k^2 \eta \right]
\end{equation}
and
\begin{equation}
\dot{\eta} = 4\pi G a^2   \sum_{i} \left(\rhob^{(i)} + \Pb^{(i)}\right) \theta^{(i)},  
\end{equation}
where $\delta \equiv \delta \rho/\rhob$ and $\eta = -\frac{1}{6} \left( k^2 \nu + h\right)$ with the label $i$ running over all fluids including the scalar field.
We now exhibit the field equations for the (coupled) quintessence field and for CDM  for the three types of cases.
Let us note that if no label is placed on either $\delta$ or $\theta$, then they are meant to refer to the CDM fluid.

\subsubsection{Type-1}
Coupled quintessence in the Type-1 case is described by the function $F = Y + V(\phi)$.
We also consider only the case for which the fluid is CDM so that $f = n e^{\alpha(\phi)}$.
The scalar field energy density and pressure for this type are given by
\begin{equation}
\rhobp =  \frac{1}{2}\frac{\dot{\phib}^2}{a^2}+V(\phi) 
\label{eq_rhophi_1_FRW}
\end{equation}
and
\begin{equation}
\Pbp =    \frac{1}{2}\frac{\dot{\phib}^2}{a^2}-V(\phi)
\label{eq_Pphi_1_FRW}
\end{equation}
for the background FRW.

The scalar field equation (\ref{eq_gen_phi}) in the case of quintessence simplifies to
\begin{equation}
 \ddot{\phib} +2 \adotoa \dot{\phib} +a^2 V_\phi = - a^2 \rhob \alpha_\phi.
\label{eq_phi_T1_FRW_quint}
\end{equation}
while the evolution of the CDM density $\rhob$ is found
from the conservation equation (\ref{eq_cons_law}) as 
\begin{equation}
 \dot{\rhob} +3 \rhob\adotoa=  \rhob \alpha_\phi \dot{\phib} 
\end{equation}
which has  a formal solution 
\begin{equation}
\rhob = \frac{\rho_0}{a^3} e^{\alpha(\phi)}.
\end{equation}
Note how this no longer falls off as conventional matter because of the coupling to $\phi$. 
The fluid equation (\ref{eq_gen_fluid_spat}) is identically satisfied.

The perturbed scalar field energy density $\rhop$ and pressure $\Pp$ are read-off from (\ref{eq_T_phi_1}) with 
$\phi = \phib + \varphi$ as
\begin{equation}
\delta \rhop =  \frac{1}{a^2}\dot{\phib} \dot{\varphi} +  V_\phi \varphi 
\label{eq_rhophi_1_pert}
\end{equation}
and
\begin{equation}
\delta \Pp =   \frac{1}{a^2} \dot{\phib}  \dot{\varphi}  - V_\phi \varphi 
\label{eq_Pphi_1_pert},
\end{equation}
while the momentum divergence for the scalar field is 
\be
\thetap = \frac{\varphi}{\dot{\phib}}.
\label{theta_p}
\ee
The scalar perturbation $\varphi$ evolves according to
\begin{equation}
\ddot{\varphi}+2\adotoa \dot{\varphi}+\left(k^2+a^2V_{\phi\phi}\right)\varphi
+\frac{1}{2}\dot{\phib}\dot{h}=-a^2\alpha_\phi \rhob \delta,
\end{equation}
while the  coupled CDM equations are found from  (\ref{eq_cons_law}) and (\ref{eq_gen_fluid_spat}) as
\begin{equation}
\dot{\delta}= -k^2\theta-\frac{1}{2}\dot{h} + \alpha_\phi \dot{\varphi}
\label{eq_fluid_delta_1}
\end{equation}
and
\begin{equation}
\dot{\theta}= -\adotoa \theta  -   \alpha_\phi \dot{\phib}\left[\theta - \thetap\right]
\label{eq_fluid_theta_1}
\end{equation}
respectively.

\subsubsection{Type 2}
We consider a coupled quintessence function of the form $F = Y + V(\phi) + n h(Z)$. The scalar field energy density and pressure are as for the Type-1 case given by (\ref{eq_rhophi_1_FRW}) and (\ref{eq_Pphi_1_FRW}) respectively.
The scalar field equation (\ref{eq_phi_2_a2}) becomes
\begin{equation}
\left(1  - \frac{\rhob \beta_Z}{1+\Zb\beta}  \right) \left(\ddot{\phib}  - \adotoa \dot{\phib}  \right)
+ 3 \adotoa \dot{\phib}
+a^2 V_\phi= 0
\end{equation}
which (assuming $1+\Zb\beta - \rhob \beta_Z \neq 0$, then allows us to find $\dot{\Zb}$ as
\begin{equation}
\dot{\Zb} = \frac{  3 \adotoa \dot{\phib}/a + a V_\phi   }{ 1  - \frac{\rhob \beta_Z}{1+\Zb\beta}}.
\end{equation}
 Thus the fluid equation (\ref{eq_phi_2_b2}) gives
\begin{equation}
 \dot{\rhob}  + 3 \adotoa \rhob  = 
 \frac{ 
    V_\phi - 3\Zb \adotoa /a   
}{1 +  \Zb \beta - \rhob \beta_Z }
 \rhob \beta_Z  \dot{\phib}.
\end{equation}

At the perturbed level (\ref{eq_rhophi_1_pert}), (\ref{eq_Pphi_1_pert}) and (\ref{theta_p}) are valid for this type of coupled model as for type-1.
We find the perturbed scalar equation from (\ref{eq_phi_2_a2}) as
\begin{eqnarray}
&&
\left[ 1   -  \frac{\rhob \beta_Z}{1+\Zb\beta} \right] \left[ \ddot{\varphi} - \adotoa \dot{\varphi} \right]
+ \left[ 3  \adotoa - \rhob  \frac{d}{dZ}\left(\frac{\beta_Z}{1+Z\beta} \right)  \dot{\bar{Z}}
\right] \dot{\varphi}
\nonumber 
\\
&&
+ ( k^2 + a^2 V_{\phi\phi}) \varphi 
+\frac{1}{2}  \dot{\phib}  \dot{h}
+\frac{a \rhob \beta_Z}{1+Z\beta}  \dot{\bar{Z}}\delta
=0.
\label{eq_phi_pert_quint_syn}
\end{eqnarray}
The perturbed CDM equations are found from (\ref{eq_x2}) and (\ref{eq_phi_2_c2}) as
\begin{eqnarray}
&&
 \dot{\delta} 
+  k^2 \theta +    \frac{1}{2} \dot{h}
 =
\frac{1}{a}\bigg\{   \frac{\Zb\beta_Z}{1+\Zb\beta}\;   ( \ddot{\varphi} - \adotoa \dot{\varphi})
\nonumber 
\\
&&
\qquad 
\ \ \ \
+  \frac{d}{dZ}\left[\frac{Z\beta_Z}{1+Z\beta}\right]   \dot{\bar{Z}}   \dot{\varphi}
\bigg\}
\label{eq_CDM_dens_pert_syn}
\end{eqnarray}
and
\begin{eqnarray}
  \dot{\theta}  + \adotoa    \theta  = \frac{1}{a^2} \beta_Z \dot{\phib} \frac{3 \adotoa \dot{\phib} + a^2 V_\phi }{ 1 + Z \beta - \rhob \beta_Z} \left[ \theta - \thetap \right].
\label{eq_CDM_mom_pert_syn}
\end{eqnarray}

\subsubsection{Type 3}
For the Type-3 case we consider a coupled quintessence function of the form $F = Y + V(\phi) + \gamma(Z)$.
In the Type-3 case with this choice of function, the scalar field energy density is given by
\begin{equation}
\rhobp =  \frac{1}{2}\frac{\dot{\phib}^2}{a^2}  +  \frac{\dot{\phib}}{a} \gamma_Z  + \gamma + V, 
\end{equation}
while the pressure is 
\begin{equation}
\Pbp =  \frac{1}{2}\frac{\dot{\phib}^2}{a^2}   - \gamma - V. 
\end{equation}
The scalar field equation becomes
\begin{equation}
\left( 1 -  \gamma_{ZZ} \right) ( \ddot{\phib} - \adotoa \dot{\phib})
+ 3a \adotoa \left( \gamma_Z - \Zb   \right) 
+a^2 V_\phi  = 0.
\end{equation}
The Type-3 case is special in that no coupling appears at the background level, regarding  the fluid equations. They remain the same as in the uncoupled case.
Furthermore, the energy-conservation equation remains uncoupled even at the linear level. In this sense, Type-3 provides for a pure momentum-transfer coupling up-to linear order
in perturbation theory.

The perturbed scalar field energy density and pressure are given by
\begin{equation}
\delta \rhop  = -\frac{1}{a}\Zb (1  - \gamma_{ZZ}) \dot{\varphi}
  +  V_{\phi} \varphi
\end{equation}
and
\begin{equation}
\delta \Pp = - \frac{1}{a} \left(\Zb - \gamma_Z\right)\dot{\varphi} - V_{\phi}\varphi
\end{equation}
respectively while its momentum divergence is
\begin{equation}
\thetap =\frac{1}{\gamma_Z - \Zb}\left(\frac{1}{a}\varphi+ \gamma_Z\theta\right).
\label{eq:thetaphiType3}
\end{equation}
This is one other notable difference from other types of coupling, namely that $\thetap$ depends also on the CDM momentum divergence $\theta$.

For the choice of $F$ above, the linearized scalar field equation is found to be
\begin{eqnarray}
&&
   (1  - \gamma_{ZZ}) (\ddot{\varphi}    + 2 \adotoa \dot{\varphi})
 - \gamma_{ZZZ} \dot{\Zb} \dot{\varphi}
+  \left(k^2 + a^2 V_{\phi\phi} \right) \varphi
\nonumber 
\\
&&
\ \ \ \
+ \frac{1}{2} (  \dot{\phib}  + a \gamma_Z) \dot{h}  
+ a k^2  \gamma_Z  \theta  
= 0,
\end{eqnarray}
while the momentum-transfer equation (\ref{eq_gen_fluid_spat}) becomes
\begin{equation}
 \dot{\theta} + \adotoa  \theta = 
 \frac{1}{a \left(\rhob  - \Zb \gamma_Z\right)} \left[ ( 3  \adotoa \gamma_Z +  \gamma_{ZZ} \dot{\Zb} )  \varphi +  \gamma_Z  \dot{\varphi} \right].
 \label{dP_type3_pert}
\end{equation}

\subsection{Results and Discussion}

\subsubsection{Type 1}

Using the derived equations for the Type 1 sub case under consideration, we use a modified version of the \texttt{CAMB} code \cite{camb}
to study the effect of the coupling to the evolution of the background as well as to the CMB temperature and matter power spectra. We have to choose a specific form for the function $\alpha(\phi)$. For simplicity we choose $\alpha(\phi) = \alpha_0 \phi$, with $a_0$ a constant.
We also have to choose a form for the quintessence potential $V(\phi)$. We will use the single exponential form (1EXP)
\be
V(\phi)=V_0 e^{-\lambda \phi},
\ee with $\lambda = 1.22$. This case has also been studied in \cite{Xia:2009zzb}. In order to compare the uncoupled $\alpha_0 = 0$ case with 
the coupled case we keep the parameter $\lambda$ of the potential fixed while $V_0$ and the CDM density $\rho_0$ are varied (for a given non-zero $\alpha_0$) 
such that each cosmology evolves to the PLANCK cosmological parameter values \cite{Ade:2013zuv}. Our initial conditions for the quintessence field are $\phi_i=10^{-4}, \dot{\phi}_i=0$. Note, however, that in this case and using the 1EXP potential, the same cosmological evolution is expected for a wide range of initial conditions.
\begin{figure}[H]
\centering
\includegraphics[scale=0.33]{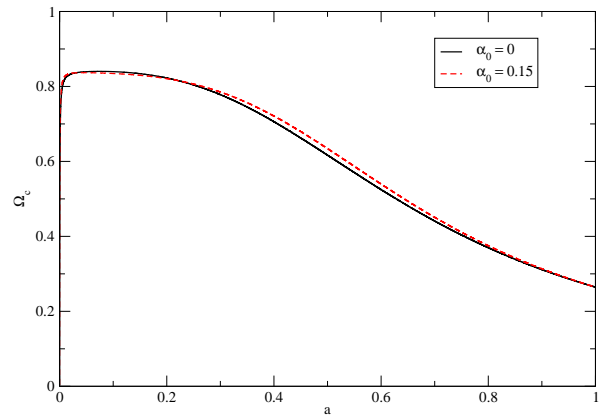}
\caption{The cosmological evolution of the CDM energy density parameter $\Omega_c$ for the Type 1 sub case under consideration. The solid black (dashed red) line is 
for $\alpha_0=0$ (0.15).}
\label{fig:densitiesType1}
\end{figure}

In Fig.~\ref{fig:densitiesType1} we show the cosmological evolution of the CDM energy density parameter $\Omega_c$. The solid black line is for $\alpha_0=0$,
 i.e. the uncoupled case, while the dashed red line is for a coupling strength $\alpha_0=0.15$.
We see that the CDM density is higher at early times for the coupled case --- this means that the matter-radiation equality will occur earlier in the 
coupled case.

\begin{figure}[H]
\centering
\includegraphics[scale=0.54]{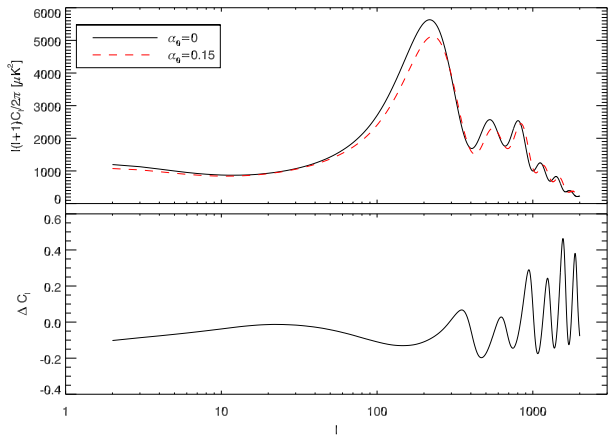}
\caption{Comparison between the Type 1 model with coupling strength $\alpha_0 = 0.15$ (dashed red line)
and the uncoupled $\alpha_0=0$ (black solid line) model. \textit{Top}: 
CMB power spectra. \textit{Bottom}: Fractional differences $\Delta C_\ell$. }
\label{fig:Type1Cls}
\end{figure}

In Fig.~\ref{fig:Type1Cls}, top panel, we show the CMB spectra $\alpha_0=0$ (uncoupled case), $\alpha_0=0.15$ (coupled case), while in the bottom panel we plot the fractional differences between the coupled and uncoupled cases 
$\Delta C_\ell = (C^{\alpha_0=0.15}_\ell-C^{\alpha_0=0}_\ell)/C^{\alpha_0=0}_\ell$.

\begin{figure}[H]
\centering
\includegraphics[scale=0.33]{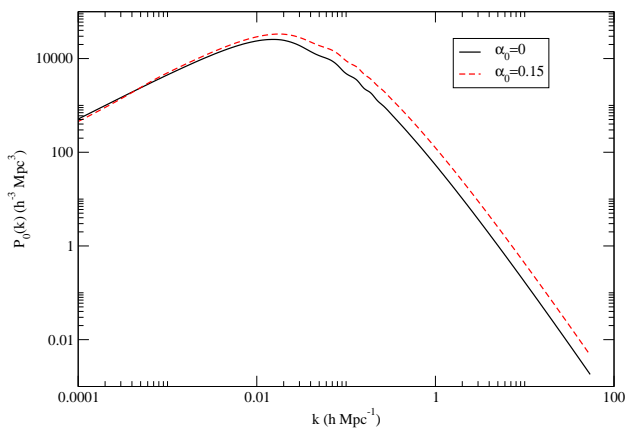}
\caption{Comparison of the linear (total)  matter power spectra at $z=0$, $P_0(k)$ between the Type 1 model with coupling strength $\alpha_0 = 0.15$ (dashed red line)
and the uncoupled $\alpha_0=0$ (solid black line) model.}
\label{fig:Type1Pk}
\end{figure}

In Fig.~\ref{fig:Type1Pk} we show the matter power spectra for $\alpha_0=0$ (uncoupled case), $\alpha_0=0.15$ (coupled case).

From Fig.~\ref{fig:Type1Pk} we see that the positive coupling affects the matter power spectrum on small scales --- small scale power is enhanced.
That is because we have an energy transfer from CDM to dark energy ($\alpha_0 \dot{\phib}<0$), 
which means that there is more dark matter at early times (see Fig.~\ref{fig:densitiesType1}), so evolving each model to the same cosmological parameters today the epoch of matter-radiation equality occurs earlier in the coupled model relative to the uncoupled one. Only small scale perturbations have time to enter the horizon and grow during the radiation dominated era, so the turnover in the matter power spectrum happens on smaller scales. The growth of perturbations in the coupled model under consideration is enhanced, small scale power is increased and the value of $\sigma_8$ is larger. 

From Fig.~\ref{fig:Type1Cls} we see that the amplitude on the small scales CMB temperature power spectrum decreases with increasing coupling. That is for the same reason as before, i.e. that the epoch of matter-radiation equality occurs earlier and the small scale anisotropies decrease. Also note that the locations of the CMB peaks have shifted towards smaller scales. This effect is strongly coupling
parameter value dependent for a given sub-case. Here, it is evident with $\alpha_0=0.15$, but it would be much smaller if we had used $\alpha_0 \sim 0.1$.
There is also a visible effect on the large-scale anisotropies (late ISW effect). For positive coupling strength the anisotropies on the very large scales ($\ell < 20$) are suppressed --- but one should bare in mind that these scales are greatly affected by cosmic variance. However, one can deal with this problem using the cross correlation between CMB temperature fluctuations and the density of galaxies (see \cite{Xia:2009zzb} for details). 

\subsubsection{Type 2}
For Type 2 we choose a sub case with 
$\beta(Z)= \frac{\beta_0}{Z}$,
where $\beta_0$ is a constant parameter so that $f \propto Z^{\beta_0/(1-\beta_0)}$. The potential $V(\phi)$ is the 1EXP potential with $\lambda=1.22$, and each cosmology evolves to the PLANCK cosmological parameter values, as before. 
The background CDM equation can then be formally solved to give
\begin{equation}
\rhob = \frac{\rho_0}{a^{3}}\Zb^{\frac{\beta_0}{1-\beta_0}}.
\label{eq:rhobctype2}
\end{equation}
We note that since $\Zb=-\dot{\phib}/a$, $\rhob_c$ depends on the time derivative $\dot{\phib}$ instead of $\phib$ itself which is a notable difference from the Type-1 case.

As before, we use the above equations for the Type 2 sub case under consideration to study the effect of the coupling to the evolution of the background as well as to the CMB temperature and matter power spectra. 
The specific form of the coupling function for the Type-2 sub case under consideration suggests that in order to have meaningful solutions, $\Zb$ has to be positive throughout the cosmological evolution.  A set of initial conditions that satisfies this constraint is $\phi_i=-0.115, \dot{\phi}_i=-10^{-4}$.

\begin{figure}[H]
\centering
\includegraphics[scale=0.33]{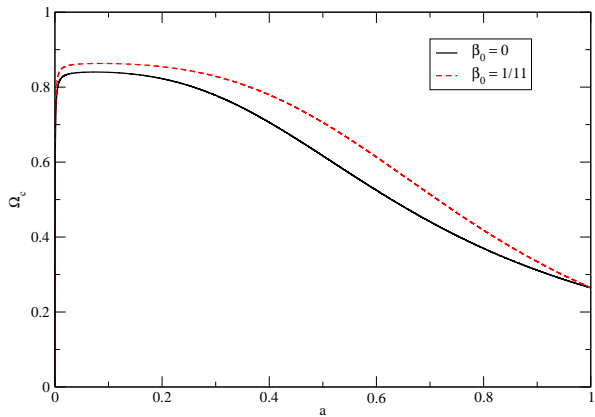}
\caption{The cosmological evolution of the CDM energy density parameter $\Omega_c$ for the Type 2 sub case under consideration. The solid black (dashed red) line is 
for $\beta_0=0$ (1/11).}
\label{fig:densitiesType2}
\end{figure}

In Fig.~\ref{fig:densitiesType2} we show the cosmological evolution of the CDM energy density parameter $\Omega_c$. The solid black line is for $\beta_0=0$,
 i.e. the uncoupled case, while the dashed red line is for a typical coupling strength $\beta_0=1/11$.
The CDM density is higher for the coupled case relative to the uncoupled one, practically at all times before converging to the same value today. Comparing with Fig.~\ref{fig:densitiesType1} for the Type 1 sub case we see that the effect is much stronger in the Type 2 model. 

\begin{figure}[H]
\centering
\includegraphics[scale=0.54]{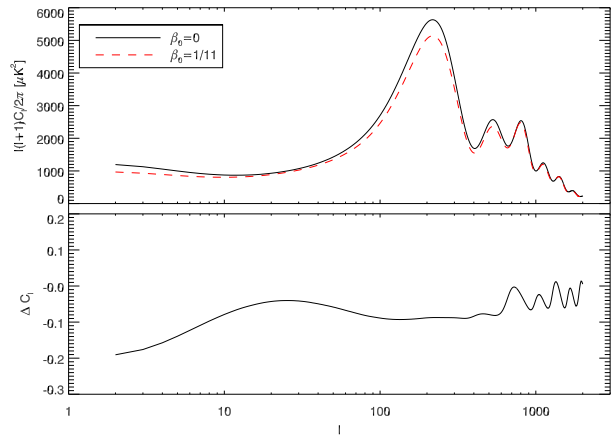}
\caption{Comparison between the Type 2 model with coupling strength $\beta_0 = 1/11$ (dashed red line)
and the uncoupled $\beta_0=0$ (solid black line) model. \textit{Top}: 
CMB power spectra. \textit{Bottom}: Fractional differences $\Delta C_\ell$. }
\label{fig:Type2Cls}
\end{figure}

\begin{figure}[H]
\centering
\includegraphics[scale=0.33]{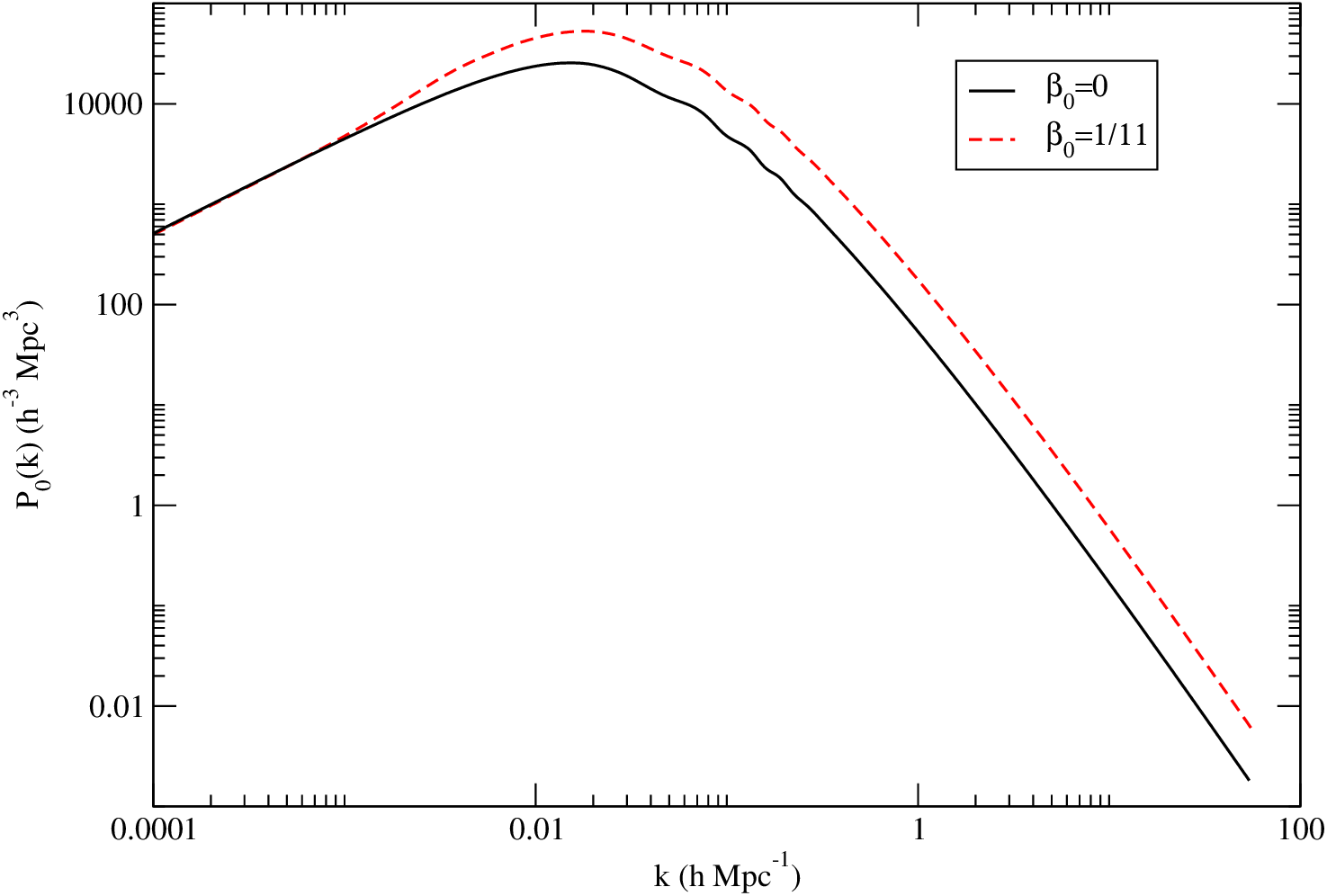}
\caption{Comparison of the linear (total)  matter power spectra at $z=0$, $P_0(k)$ between the Type 1 model with coupling strength $\beta_0 = 1/11$ (dashed red line)
and the uncoupled $\beta_0=0$ (solid black line) model.}
\label{fig:Type2Pk}
\end{figure}

In Fig.~\ref{fig:Type2Cls} and Fig.~\ref{fig:Type2Pk} we show the CMB and matter power spectra, respectively for $\beta_0=0$ (uncoupled case), $\beta_0=1/11$ (coupled case).  
From Fig.~\ref{fig:Type2Pk} we see that the small scale power (and, consequently, the value of $\sigma_8$) is larger for the coupled case, as expected from the evolution of the CDM energy density. 
From Fig.~\ref{fig:Type2Cls} 
we see that the suppression of the large-scale anisotropies which arise from the ISW effect is significant, as is the suppression of the first peak magnitude. One of the effects of the coupling on the (small-scale) CMB spectrum is that the location of
the acoustic peaks is shifted towards larger $\ell$. While this effect was evident in the Type-1 case (see Fig.~\ref{fig:Type1Cls}), here it is negligible. As we have already mentioned, this effect strongly depends on the value of the coupling parameter. 
Moreover, the location of the peaks is also affected by other cosmological factors. The position of the first peak, for example, is characterized by the shift parameter, which depends on the size of the sound horizon at decoupling. This depends on the behavior of the Hubble length $1/{\cal H}(z)$, which in turn depends on the evolution of the quintessence field.
In general, the small scale effects on the CMB power spectrum have a non-trivial dependence on the cosmology of each model, and thus are sensitive to both the coupling strength and the evolution of the quintessence field.

\subsubsection{Type 3}

For Type 3 we choose $\gamma(Z) = \gamma_0 Z^2$ for simplicity. Defining a new type of metric as $\tilde{g}^{\mu\nu} = g^{\mu\nu} + 2 \gamma_0 u^\mu u^\nu$,
this function transforms the scalar field action into
\bea
 \nonumber
S_{\phi}&=&
-\int d^4x \sqrt{-g}\left[
   \frac{1}{2}\tilde{g}^{\mu \nu} \phi_\mu \phi_\nu
+ V(\phi)
\right]
\\ \nonumber
&\rightarrow & 
\int dt \, d^3x \, a^3 \left[
    \frac{1}{2}(1 - 2 \gamma_0)  \dot{\phi}^2
 - \frac{1}{2} |\grad\phi|^2
- V(\phi)
\right]
\eea
where the arrow denotes working in a frame where the CDM $3$-velocity is zero.
Hence, the model is physically acceptable for $\gamma_0 < \frac{1}{2}$. For $\gamma_0 \rightarrow 1/2$
we have a strong coupling problem, while for $\gamma_0>1/2$ the kinetic term is negative, so there is an associated ghost. Introducing the coupling at the level of the action has therefore helped us identify pathologies.

In Type 3 theories, the background densities evolve as in the uncoupled case. For the specific sub case under consideration we choose $\gamma_0=0.15$ and our initial conditions for the quintessence field are the same as in the Type-1 sub case, i.e.
$\phi_i=10^{-4}, \dot{\phi}_i=0$. The potential $V(\phi)$ is the 1EXP potential with $\lambda=1.22$, and each cosmology evolves to the PLANCK cosmological parameter values, as before. 

\begin{figure}[H]
\centering
\includegraphics[scale=0.33]{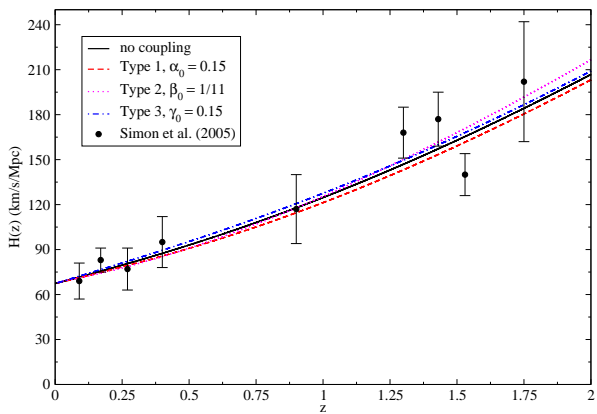}
\caption{The Hubble parameter versus redshift for the uncoupled model and the three non-zero coupling sub cases, together with expansion history measurements.}
\label{fig:Hz}
\end{figure}

In Fig.~\ref{fig:Hz} we show the evolution of the Hubble parameter $H(z)$ for all three sub cases and the uncoupled model, together with the expansion history measurements from Simon et al. (2005) \cite{Simon:2004tf}.

\begin{figure}[H]
\centering
\includegraphics[scale=0.54]{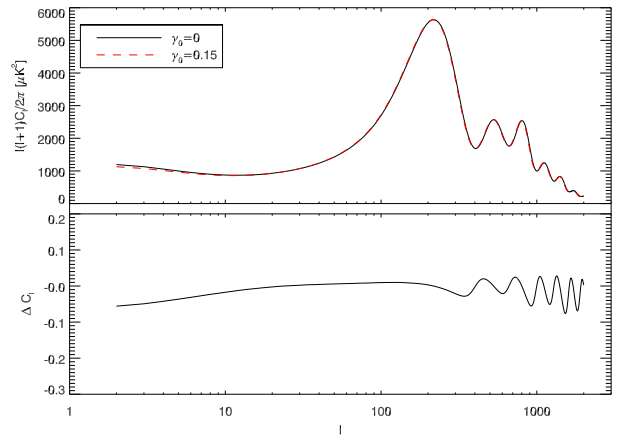}
\caption{Comparison between the Type 3 model with coupling strength $\gamma_0 = 0.15$ (red dashed line)
and the uncoupled $\gamma_0=0$ (solid black line) model. \textit{Top}: 
CMB power spectra. \textit{Bottom}: Fractional differences $\Delta C_\ell$. }
\label{fig:Type3Cls}
\end{figure}

\begin{figure}[H]
\centering
\includegraphics[scale=0.33]{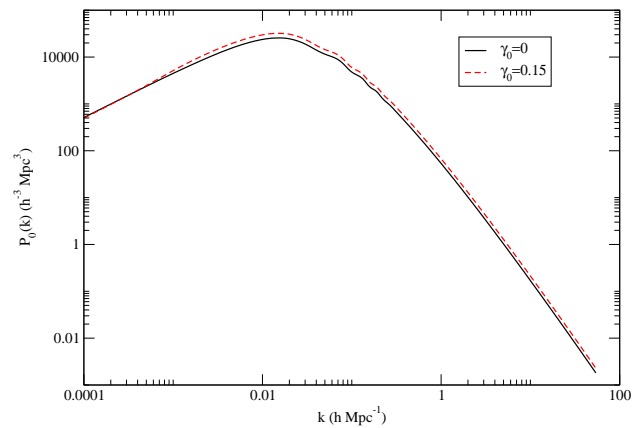}
\caption{Comparison of the linear (total)  matter power spectra at $z=0$, $P_0(k)$ between the Type 3 model with coupling strength $\gamma_0 = 0.15$ (red dashed line)
and the uncoupled $\gamma_0=0$ (solid black line) model.}
\label{fig:Type3Pk}
\end{figure}

 In Fig.~\ref{fig:Type3Cls} and Fig.~\ref{fig:Type3Pk} we show the CMB and matter power spectra, respectively, for $\gamma_0=0$ (uncoupled case), $\gamma_0=0.15$ (coupled case). We see that the relative effect of the Type 3 coupling is minute in comparison with Type 1 and Type 2 models. This is not surprising, as the Type-3 case is unique in the sense that there is no coupling at the background level and the evolution of the CDM density contrast is the same as in the uncoupled case --- it is a pure momentum-transfer coupling.

\section{Conclusions}

We have presented a novel method to construct general interacting dark energy models, which is inspired by the pull-back formalism for relativistic fluids. By introducing the coupling at the level of the action, instead of the usual phenomenological choice at the level of the field equations, we have shown that the existing models in the literature are only specific sub cases that belong in general \emph{classes} of coupled dark energy theories. An advantage of introducing the coupling at the level of the action is that pathologies (like ghosts and strong coupling problems) can be easily identified. 

Using our formalism we have constructed three distinct general classes of models of dark energy in the form of a scalar field which is then explicitly coupled to dark matter. We then studied specific sub cases by choosing a form for the quintessence potential and the coupling function.  Our choices for the coupling were specifically chosen so that we could clearly demonstrate its effect in the expansion history of the Universe, as well as in the CMB and matter power spectra. For the Type-1 sub case we considered, the value of the coupling constant $\alpha_0$ is already constrained to be $< 0.1$ (see, for example, \cite{Amendola:1999er, Xia:2009zzb}). We plan to present constraints for the Type-2 and Type-3 sub cases in a future publication.
Observationally, the most interesting theory seems to be Type-3, which appears to be able to mimic the uncoupled case for a wider range of coupling values than the other two Types. 

However, we should keep in mind that these models depend on a few free parameters, one of which is the form of the quintessence potential $V(\phi)$. We have studied the same sub cases using the double exponential (2EXP) potential. For the Type-1 sub case, the effect of the positive coupling constant on the CMB and matter power spectra is very similar to the one using the 1EXP potential on small scales, but we see a slight increase on the late-ISW effect, instead of a suppression. The Type-2 behavior using the 2EXP potential is very different than the one using the 1EXP, since the former can very well mimic the uncoupled case up to an order $0.1$ coupling. Type 3 exhibits very similar behavior for both potentials, but when we increase the coupling further than, say, $0.3$, the ISW effect strongly deviates from the uncoupled case. In general, there is a wide parameter space that must be investigated in order to constrain or rule out specific models. 

Clearly the observational spectra depend on the type of coupling model but also on the specific potential and specific coupling function used. It would then seem rather futile to try and distinguish 
these models from each other. However, as we will show in a future work, it is possible to parameterize the field equations by introducing all possible types of terms that can appear in a coupled
theory akin to the Parameterized Post-Friedmannian approach of~\cite{Skordis:2008vt,Baker:2011jy,Baker:2012zs}. This would pave the way for a completely model-independent approach to
testing models of coupled dark energy by reducing the large parameter space of free functions and coupling types to a small set of constants. 

Our theories may be able offer a viable alternative to $\Lambda$CDM. To confirm this a more detailed theoretical and numerical analysis is required. We plan to present such an analysis in a future work.

\section{Acknowledgments}
E.J.C. acknowledges support form the Royal Society, STFC and the Leverhulme Trust. C.S. acknowledges support from the Royal Society. A.P.'s research is supported by the project GLENCO, funded under
the FP7, Ideas, Grant Agreement n. 259349. A.P. also received support from the STFC (UK) while part of this work was in progress at the University of Manchester. A.P. acknowledges the University of Nottingham for hospitality during various stages of this work. The authors would like to thank Richard Battye, Jonathan Pearson and Fergus Simpson for useful discussions, and Ewan Tarrant for help with numerics.

\bibliographystyle{apsrev}
\bibliography{references}

\end{document}